\documentclass[aps,prc,superscriptaddress,showpacs,floatfix,nofootinbib,notitlepage,twocolumn]{revtex4-1}
\usepackage{amsmath,graphicx,float,hyperref,upgreek}
\newcommand{\trento}{T$\mathrel{\protect\raisebox{-2.1pt}{R}}$ENTo}

\begin{document}

\title{Correlation between mean transverse momentum and anisotropic flow \\ in heavy-ion collisions}

\author{Giuliano Giacalone}
\affiliation{Universit\'e Paris Saclay, CNRS, CEA, Institut de physique th\'eorique, 91191 Gif-sur-Yvette, France}
\author{Fernando G. Gardim}
\affiliation{Instituto de Ci\^encia e Tecnologia, Universidade Federal de Alfenas, 37715-400 Po\c cos de Caldas, MG, Brazil}
\author{Jacquelyn Noronha-Hostler}
\affiliation{Department of Physics, University of Illinois at Urbana-Champaign, Urbana, IL 61801, USA}
\author{Jean-Yves Ollitrault}
\affiliation{Universit\'e Paris Saclay, CNRS, CEA, Institut de physique th\'eorique, 91191 Gif-sur-Yvette, France}

\begin{abstract}
  The correlation between the mean transverse momentum of outgoing particles, $\langle p_t \rangle$, and the magnitude of anisotropic flow, $v_n$, has recently been measured in Pb+Pb collisions at the CERN Large Hadron Collider, as a function of the collision centrality.
  We confirm the previous observation that event-by-event hydrodynamics predicts a correlation between $v_n$ and $\langle p_t \rangle$ that is similar to that measured in data.
  We show that the magnitude of this correlation can be directly predicted from the initial condition of the hydrodynamic calculation, for $n=2,3$, if one replaces $v_n$ by the corresponding initial-state anisotropy, $\varepsilon_n$, and $\langle p_t\rangle$ by the total energy per unit rapidity of the fluid at the beginning of the hydrodynamic expansion. 
\end{abstract}
\date{\today}

\maketitle

\section{Introduction}

Anisotropic flow has been much studied in ultrarelativistic heavy-ion collisions, as it probes the properties of the little quark-gluon plasma formed in these collisions~\cite{Heinz:2013th}.
The event-by-event fluctuations of $v_n$, the $n^{\rm th}$ Fourier harmonic of the azimuthal distribution of the emitted hadrons, have been precisely characterized~\cite{Aad:2013xma,Sirunyan:2017fts,Acharya:2019vdf}, as well as the mutual correlations between different flow harmonics~\cite{Aad:2014fla,ALICE:2016kpq,Sirunyan:2019izh,Acharya:2020taj}.
Recently, following a suggestion by Bo\.zek~\cite{Bozek:2016yoj}, the ATLAS Collaboration has measured a correlation of a new type, namely, the correlation between the mean transverse momentum, $\langle p_t\rangle$, and $v_n^2$~\cite{Aad:2019fgl}, a quantity dubbed  $\rho_n$.
It is the first time that anisotropic flow, which is dimensionless, is correlated with a dimensionful quantity. 
This opens a new window on the study of $v_n$, by giving new information about its correlation with the size and/or temperature of the quark-gluon plasma. 
Although event-by-event hydrodynamic results on $\rho_n$ are in fair agreement with experimental data~\cite{Bozek:2020drh}, a clear picture of the physical mechanism that produces this correlation is still missing.
In this paper, we explain the origin of the correlation between $\langle p_t\rangle$ and $v_n$ in hydrodynamic calculations.
We first confirm, in Sec.~\ref{s:hydro}, that state-of-the-art hydrodynamic calculations yield results on $\rho_n$ that are in agreement with recent Pb+Pb data. 
We use results from hydrodynamic calculations~\cite{Giacalone:2017dud} obtained prior to the ATLAS analysis, so that they can be considered as predictions. 
We then unravel, in Sec.~\ref{s:III}, the physical mechanism behind $\rho_n$.
While it is well established that $v_2$ and $v_3$ originate on an event-by-event basis from the anisotropies of the initial density profile, $\varepsilon_n$~\cite{Alver:2006wh,Alver:2010gr,Teaney:2010vd,Gardim:2011xv,Niemi:2012aj,Giacalone:2017uqx}, the new crucial feature that we shall elucidate here is the origin of $\langle p_t\rangle$ fluctuations in hydrodynamics.
These fluctuations are thought to be driven by fluctuations in the fireball size, $R$~\cite{Broniowski:2009fm,Bozek:2012fw}, a phenomenon referred to as size-flow transmutation~\cite{Bozek:2017elk}.
We show in Sec.~\ref{s:ei} that, in fact, a much better predictor for $\langle p_t\rangle$ is provided by the initial total energy per rapidity of the fluid~\cite{Gardim:2020sma}.
On this basis, in Sec.~\ref{s:trento} we evaluate $\rho_n$ using a standard initial-state model for Pb+Pb collisions, and we obtain results that are in good agreement with ATLAS data. 
We further show that agreement with data is instead lost if one uses $R$ as an event-by-event predictor of $\langle p_t\rangle$.

\section{Results from event-by-event hydrodynamics}
\label{s:hydro}

The ATLAS Collaboration measured the Pearson correlation coefficient between the mean transverse momentum and the anisotropic flow of the event~\cite{Aad:2019fgl}.
Experimentally, this is obtained from a three-particle correlation introduced by Bo\.zek~\cite{Bozek:2016yoj}. 
Since self correlations are subtracted in the measure of the correlation coefficient (i.e., one does not correlate a particle with itself), this observable is insensitive to trivial statistical fluctuations and probes genuine \textit{dynamical} fluctuations~\cite{Adams:2005ka}, due to correlations.
In hydrodynamics, $\rho_n$ can be evaluated as:
\begin{equation}
  \label{rhonhydro}
  \rho_n\equiv \frac{\left\langle\langle p_t\rangle
    v_n^2\right\rangle
    -\left\langle\langle p_t\rangle\right\rangle\left\langle v_n^2\right\rangle}{\sigma_{p_t}\sigma_{v_n^2} },
\end{equation}
where, following the notation of Ref.\cite{Bozek:2012fw}, $\langle p_t\rangle$ 
denotes an average over the single-particle momentum distribution, $f(p)$, at freeze-out in a given event\footnote{Using this notation, one can also write $v_n\equiv\left|\langle e^{in\varphi}\rangle\right|$, where $\varphi$ is the azimuthal angle of the particle momentum.}, and the outer angular brackets denote an average over events in a given multiplicity (centrality) window.
$\sigma_{p_t}$ and $\sigma_{v_n}$ 
denote, respectively, the standard deviation of $\langle p_t\rangle$ and of $v_n^2$:
\begin{align}
\label{sigmapt}
\nonumber \sigma_{p_t}&\equiv \sqrt{\left\langle \langle p_t\rangle^2\right\rangle
-\left\langle \langle p_t\rangle\right\rangle^2}, \\
\sigma_{v_n^2}&\equiv \sqrt{\left\langle v_n^4\right\rangle
-\left\langle v_n^2\right\rangle^2}.
\end{align}

\begin{figure}[t!]
    \centering
    \includegraphics[width=.8\linewidth]{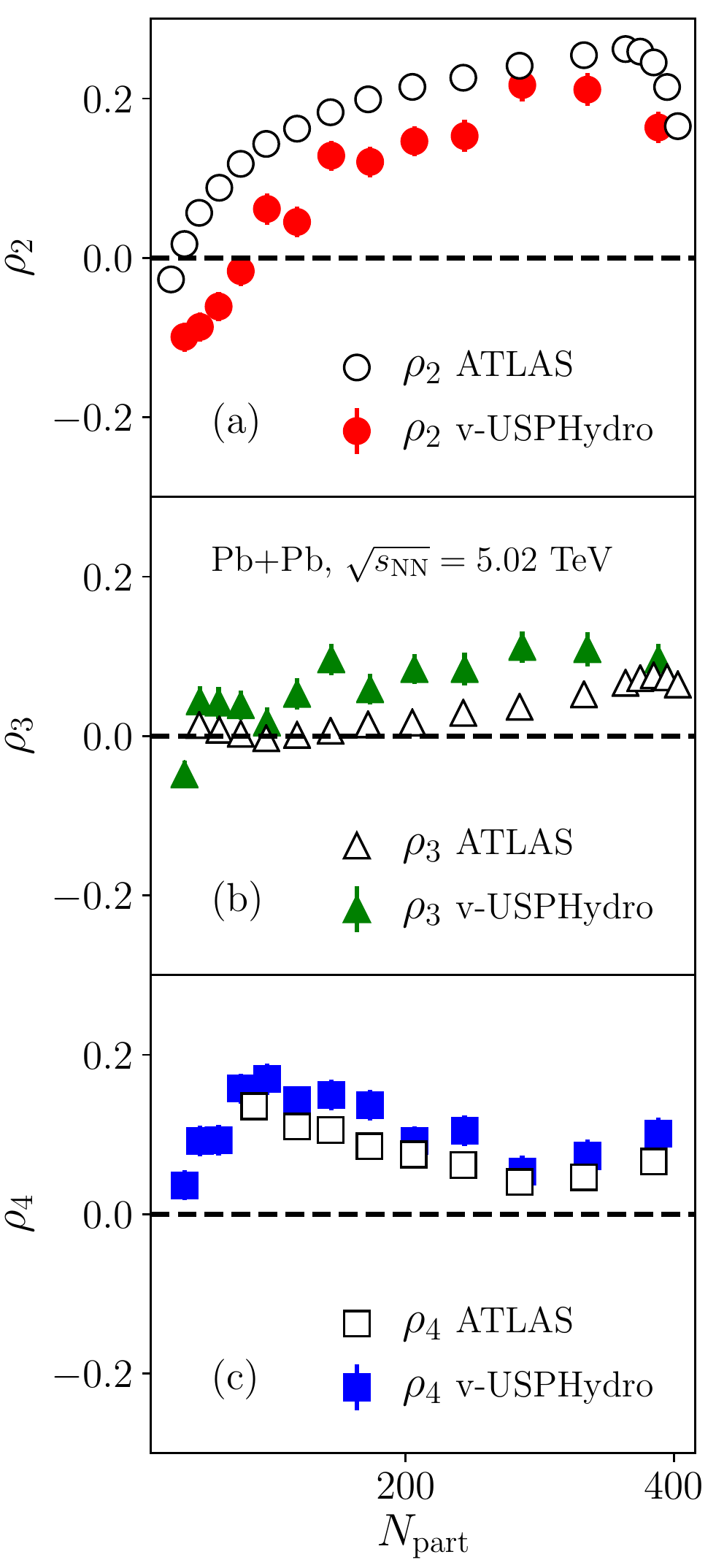}
    \caption{(Color online) Value of $\rho_n$ for $n=2$ (a), $n=3$ (b), $n=4$ (c), as a function of the number of participant nucleons in Pb+Pb collisions at $\sqrt{s_{\rm NN}}=5.02$~TeV.
      Empty symbols are experimental results from the ATLAS Collaboration~\cite{Aad:2019fgl}, integrated over $0.5<p_t<2$~GeV/c, while full symbols are hydrodynamic results~\cite{Giacalone:2017dud}, integrated over $0.2<p_t<3$~GeV/c.
 }
    \label{fig:hydro}
\end{figure}
We evaluate $\rho_n$ using hydrodynamic simulations. The setup of our calculation is the same as in Ref.~\cite{Giacalone:2017dud}. We evolve 50000~minimum bias Pb+Pb collisions at $\sqrt{s_{\rm NN}}=5.02$~TeV through the 2+1 viscous 
relativistic hydrodynamical code {\footnotesize V-USPHYDRO}
\cite{Noronha-Hostler:2013gga,Noronha-Hostler:2014dqa,Noronha-Hostler:2015coa}.
The initial condition of the evolution is the profile of entropy density generated using the \trento{} model~\cite{Moreland:2014oya}, tuned as in Ref.~\cite{Bernhard:2016tnd}.\footnote{That is, we implement a geometric average of nuclear thickness functions (parameter $p=0$), where the thickness function of a nucleus is given by a linear superimposition of the thicknesses of the corresponding participant nucleons, modeled as Gaussian density profiles of width $w=0.51$~fm. The thickness of each participant nucleon is further allowed to fluctuate in normalization according to a gamma distribution of unit mean and standard deviation $1/\sqrt{k}$, with $k=1.6$.}
%The fluid was assumed to possess longitudinal boost invariance~\cite{Bjorken:1982qr}.
We neglect the expansion of the system during the pre-equilibrium phase~\cite{Vredevoogd:2008id,vanderSchee:2013pia,Kurkela:2018wud}, and start hydrodynamics at time $\tau_0=0.6$~fm/c after 
the collision~\cite{Kolb:2000fha}.
We use an equation of state based on lattice QCD~\cite{Borsanyi:2013bia}, and we implement a constant shear viscosity over entropy ratio $\eta/s=0.047$~\cite{Alba:2017hhe}.
Fluid cells are transformed into hadrons~\cite{Teaney:2003kp} when the local temperature drops below $150$~MeV. All hadronic resonances can be formed during this freeze-out process, and we implement subsequent strong decays into stable hadrons.
To mimic the centrality selection performed in experiment, we sort events into centrality classes according to their initial entropy (5\% classes are used).
We evaluate hadron observables by integrating over the transverse momentum range $0.2<p_t<3$~GeV/c, and over $|\eta|<0.8$.

In Fig.~\ref{fig:hydro} we show our results along with ATLAS data.
The kinematic cuts in $p_t$ and $\eta$ in our hydrodynamic calculation differ from those of ATLAS. 
The reason is that we use results from a prior high-statistics hydrodynamic calculations~\cite{Giacalone:2017dud}. 
ATLAS shows three sets of results, with two lower cuts ($0.5$ and $1$~GeV/c) and two upper cuts ($2$ and $5$~GeV/c) in $p_t$.
Since hydrodynamics is meant to explain the production of particles at low-$p_t$, we choose the lowest values for both cuts, that is, data integrated over $0.5<p_t<2$~GeV/c. 
The calculation is in reasonable agreement with data, and we conclude that event-by-event relativistic hydrodynamics captures semi-quantitatively the magnitude and the centrality dependence of $\rho_2$, $\rho_3$ and $\rho_4$.

\section{Physical origin of $\boldsymbol{\rho_2}$ and $\boldsymbol{\rho_3}$}
\label{s:III}

\subsection{Initial energy as a predictor for $\boldsymbol{\langle p_t\rangle}$}
\label{s:ei}

\begin{figure*}[t!]
    \centering
    \includegraphics[width=\linewidth]{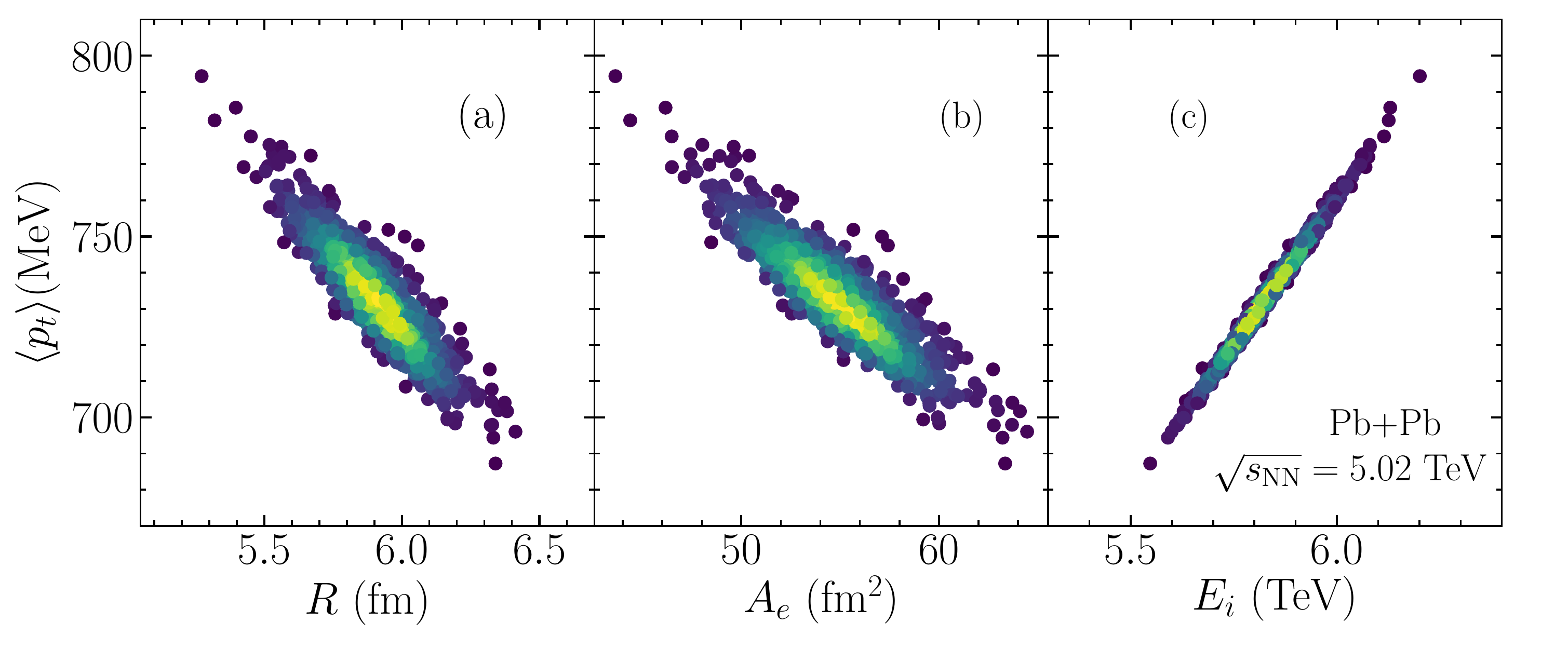}
    \caption{(Color online) Results from ideal hydrodynamic simulations of Pb+Pb collisions at $\sqrt{s_{\rm NN}}=5.02$~TeV at impact parameter $b=2.5$~fm. 850 events have been simulated, where each event has the same total entropy, but a different entropy density profile. Each symbol represents a different event.
      (a) Scatter plot of the mean transverse momentum of charged particles, $\langle p_t\rangle$, versus the initial size, $R$, defined by Eq.~(\ref{defr}).
      (b) Scatter plot  $\langle p_t\rangle$, versus the elliptic area $A_e$, defined by Eq.~(\ref{defA}).   
      (c) Scatter plot of $\langle p_t\rangle$ versus the initial energy per unit rapidity, $E_i$, defined by Eq.~(\ref{defei}). 
    }
    \label{fig:ptpredictor}
\end{figure*}

The full hydrodynamic calculation allows us to reproduce the experimental data, but it does not give much insight into the physics underlying the observed $\rho_n$. 
In this Section, we trace the origin of $\rho_n$ back to the initial state of the hydrodynamic calculation. 
It is well known that $v_2$ and $v_3$ are driven by the initial spatial eccentricity, $\varepsilon_2$, and triangularity, $\varepsilon_3$, on an event-by-event basis. 
The other ingredient entering $\rho_n$ is the mean transverse momentum of the event. 
Therefore, we must identify the property of the initial state which drives the mean transverse momentum on an event-by-event basis. 

It has recently been shown~\cite{Gardim:2020sma} that if one fixes the total entropy (which in an experiment amounts to fixing the centrality of the collision), then $\langle p_t \rangle$ is essentially determined by the energy of the fluid per unit rapidity at the initial time $\tau_0$, which we denote by $E_i$:
\begin{equation}
  \label{defei}
E_i\equiv \tau_0\int \epsilon(\tau_0,x,y)dxdy,
\end{equation}
where $\epsilon$ is the energy density, and the integral runs over the transverse plane. 
This is at variance with the earlier claims~\cite{Broniowski:2009fm} that $\langle p_t\rangle$ is determined by the initial transverse size of the fireball, $R$, defined as~\cite{Bozek:2012fw,Bozek:2017elk}:
\begin{equation}
  \label{defr}
R^2\equiv 2\frac{\int (x^2+y^2) s(\tau_0,x,y)dxdy}{\int s(\tau_0,x,y)dxdy}
\end{equation}
where $s$ is the entropy density.\footnote{The factor $2$ ensures that for a uniform entropy density $s(\tau_0,x,y)$ in a circle of radius $R$, the right-hand side gives $R^2$.}
Schenke, Shen and Teaney~\cite{Schenke:2020uqq} have recently proposed an improved predictor of $\langle p_t\rangle$, relative to (\ref{defr}).
Their idea is that the relevant quantity is the area of the overlap region, and that the elliptical shape must be taken into account in evaluating this area, which is defined as Eq.~(34) of~\cite{Schenke:2020uqq}:
\begin{equation}
  \label{defA}
A_e\equiv \frac{\pi}{2} R^2\sqrt{1-\varepsilon_2^2},
\end{equation}
where $\varepsilon_2$ is the initial eccentricity. 

To illustrate the difference between these three predictors, we have evaluated $E_i$, $R$, $A_e$ and $\langle p_t\rangle$ in event-by-event hydrodynamics at fixed initial entropy. 
Note that the minimum bias calculation performed in Sec.~\ref{s:hydro} is not well suited for this purpose, as, even if we narrowed down the width of our centrality bins by a factor 2, there would still be significant entropy fluctuations in our sample.
For this reason, we resort to the calculation shown in Ref.~\cite{Gardim:2020sma}. Here the events are evaluated at fixed impact parameter $b=2.5$~fm, and at fixed total entropy corresponding to the mean entropy of Pb+Pb collisions at $\sqrt{s_{NN}}=5.02$~TeV in the 0-5\% centrality window. Also, we perform an ideal hydrodynamic expansion, which ensures conservation of entropy. The initial condition of the calculation, the equation of state, and the initialization time, $\tau_0$, are the same used in the calculation of Sec.~\ref{s:hydro}.\footnote{The sole differences are that we implement slightly smaller entropy fluctuations, using $k=2$ in \trento{}, which provide a better fit of LHC data~\cite{Giacalone:2017dud}, as well as a slightly higher freeze-out temperature $T_f=156.5$~MeV, which has now become a more standard choice~\cite{Bazavov:2018mes}.}
This calculation is evolved through the MUSIC hydrodynamic code~\cite{Schenke:2010nt,Schenke:2010rr,Paquet:2015lta}.

Figure~\ref{fig:ptpredictor}(a) displays the scatter plot of $R$~vs.~$\langle p_t\rangle$ obtained in this calculation.
These two quantities are negatively correlated, as already shown by other authors~\cite{Bozek:2017elk}. 
The explanation is that for a fixed total entropy, a smaller size generally implies a larger entropy density, hence a larger temperature, which in turn implies a larger $\langle p_t\rangle$~\cite{Gardim:2019xjs}. 
There is, however, a significant spread of the values of $\langle p_t\rangle$ for a fixed $R$. 
A similar spread is observed if one correlates  $\langle p_t\rangle$ with the elliptic area, $A_e$, as shown in Fig.~\ref{fig:ptpredictor}(b).\footnote{Note that we consider fairly central collisions where $\varepsilon_2$ is small. This explains the similar correlation patterns in Figs. ~\ref{fig:ptpredictor}(a) and ~\ref{fig:ptpredictor}(b). The situation would likely be different in mid-central collisions where $\varepsilon_2$ is larger.}
By contrast, there is an almost one-to-one correspondence between $\langle p_t\rangle$ and the initial energy, $E_i$, as displayed in Fig.~\ref{fig:ptpredictor}~(b). %\footnote{Interestingly, the correlation of $\langle p_t \rangle$ with the {\it final\/} energy evaluated at freeze-out is not as strong as the correlation with $E_i$~\cite{Gardim:2020sma}, and we do not have a simple explanation for this observation.} 
We shall show in Sec.~\ref{s:trento} that, in order to understand the measured correlation between $\langle p_t \rangle$ and $v_n$, it is indeed crucial to employ $E_i$ as a predictor of the average transverse momentum.

%Our study explores a different direction, where one uses the information from the initial energy, not just from the initial geometry. 

\subsection{Results from models of initial conditions} 

\label{s:trento}

\begin{figure*}[t!]
    \centering
    \includegraphics[width=.95\linewidth]{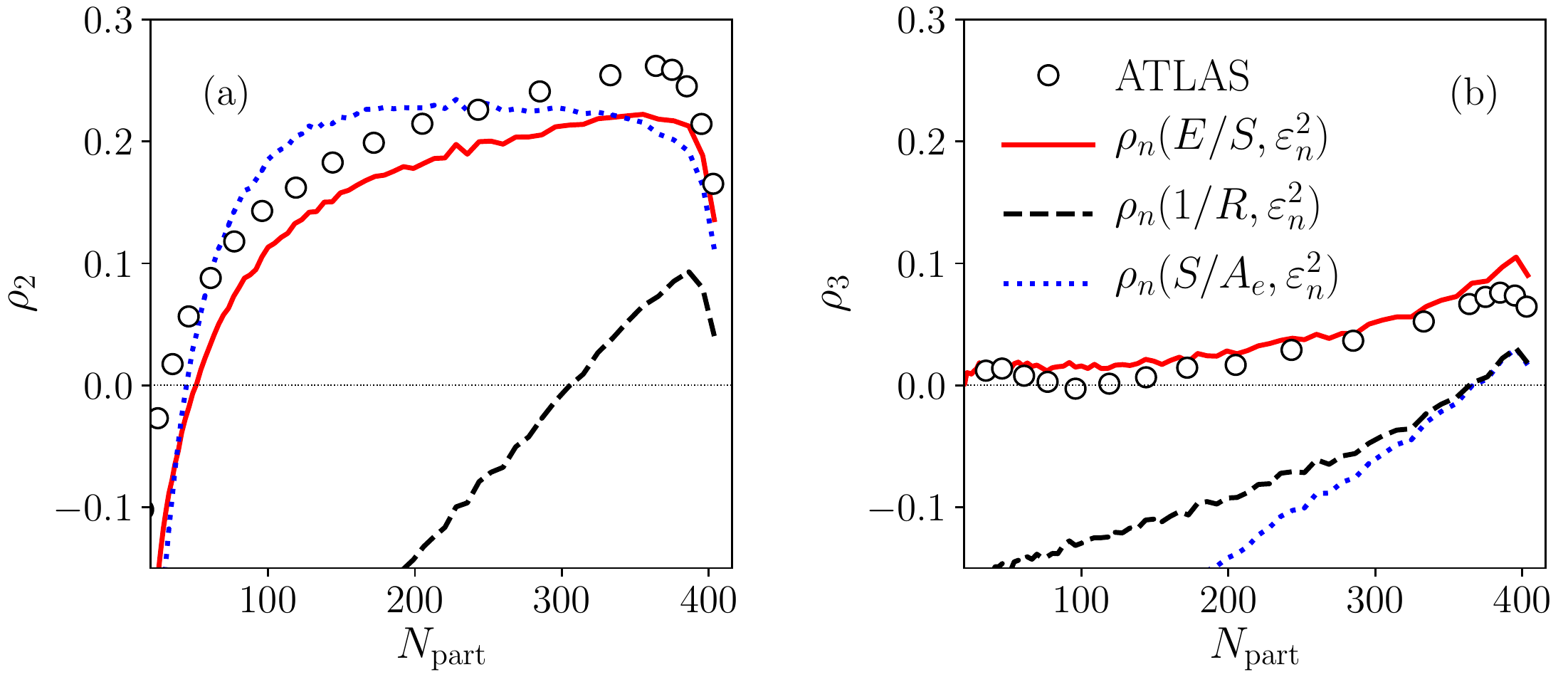}
    \caption{(Color online) Variation of $\rho_2$ (a) and $\rho_3$ (b) with the number of participants in Pb+Pb collisions at $\sqrt{s_{\rm NN}}=5.02$~TeV.
      As in Fig.~\ref{fig:hydro}, symbols are experimental results from the ATLAS Collaboration~\cite{Aad:2019fgl}.
      The shaded band is our result using Eq.~(\ref{rhoninitial}). The width of the band is the statistical error evaluated through jackknife resampling. The dashed lines are obtained by replacing $E_i/S$ with $1/R$ in Eq.~(\ref{rhoninitial}), and the dotted lines by replacing $E_i/S$ with $S/A_e$.      }
    \label{fig:trento}
\end{figure*}

We first explain how $\rho_n$, defined in Eq.~(\ref{rhonhydro}), can be evaluated from the initial conditions of the hydrodynamic calculations.
First, one uses the approximate proportionality between $v_n$ and the initial anisotropy $\varepsilon_n$~\cite{Niemi:2012aj} for $n=2,3$:
\begin{equation}
\label{linearresponse}
v_n=\kappa_n\varepsilon_n,   
\end{equation}
where $\kappa_n$ is a response coefficient which is the same for all events at a given centrality.\footnote{This linear response works for $v_2$ and $v_3$, but not for $v_4$~\cite{Gardim:2011xv}, so that we do not discuss $\rho_4$ in this section.}

Next, one uses the observation, made in Sec.~\ref{s:ei}, that $\langle p_t\rangle$ is a function of the initial energy $E_i$ for a fixed total entropy. 
We first relax the stringent condition that the total entropy $S$ is fixed. 
Since $\langle p_t\rangle$ is the average transverse momentum {\it per particle\/}, the relevant predictor is also the initial energy {\it per particle\/}, which is proportional to $E_i/S$ since the entropy is proportional to the number of particles. 
We therefore replace $E_i$ with $E_i/S$ from now on,\footnote{In practice, the results presented below use very narrow entropy (centrality) bins, so that one would obtain the same results using $E_i$ or $E_i/S$. If one uses $E_i/S$, results are unchanged if one uses wider centrality bins, up to 2\%.  A moderate variation starts to be visible if one uses 5\% bins, as with other observables~\cite{Gardim:2016nrr}.}and we assume that $\langle p_t\rangle=f(E_i/S)$, where $f(E_i/S)$ is some smooth function of  $E_i/S$.

Linearizing in the fluctuations of $E_i/S$ and $\langle p_t\rangle$ around their mean values, $\langle E_i/S\rangle$ and $\left\langle\langle p_t\rangle\right\rangle$, one obtains
\begin{equation}
  \label{linearized}
  \langle p_t\rangle -\left\langle\langle p_t\rangle\right\rangle=
  f'(\langle E_i/S\rangle) \left(\frac{E_i}{S}-\left\langle\frac{E_i}{S}\right\rangle\right). 
\end{equation}
Inserting Eqs.~(\ref{linearresponse}) and (\ref{linearized}) into  Eq.~(\ref{rhonhydro}), one obtains:
\begin{equation}
  \label{rhoninitial}
  \rho_n= \frac{\left\langle\frac{E_i}{S}
    \varepsilon_n^2\right\rangle
    -\left\langle \frac{E_i}{S}\right\rangle\left\langle \varepsilon_n^2\right\rangle}{
    \sigma_{E_i/S}    \sigma_{\varepsilon_n^2}
}\frac{f'(\langle E_i/S\rangle)}{\left| f'(\langle E_i/S\rangle)\right|}
    ,
\end{equation}
where $\sigma_{E_i/S}$ and $\sigma_{\varepsilon_n^2}$ denote the standard deviations, obtained by replacing $\langle p_t\rangle$ and $v_n$ with $E_i/S$ and $\varepsilon_n$ in Eq.~(\ref{sigmapt}). 
Remarkably enough, the dependence on the unknown function $f(E_i/S)$ cancels, except for the sign of $f'(\langle E_i/S\rangle)$. 
An important advantage of Eq.~(\ref{rhoninitial}) is that it allows us to evaluate $\rho_n$ in millions of simulated initial conditions with little computational effort.
%and thus to overcome the issue of large entropy fluctuations within a finite centrality bin~\cite{Bozek:2020drh}. This allows us, hence, to evaluate $\rho_n$ in the strict limit of fixed initial entropy, and to reproduce the situation of Fig.~\ref{fig:ptpredictor} in a minimum bias calculation.
We have generated 20 million minimum bias Pb+Pb events using the same \trento{} parametrization as in Fig.~\ref{fig:ptpredictor}. We sort the events into narrow $0.25\%$ centrality bins, and in each bin we evaluate $\rho_n$ according to Eq.~(\ref{rhoninitial}). 
 To evaluate $E_i$ in each event, we assume that the entropy profile returned by \trento{}, $s$, is related to the energy density, $\epsilon$, of the event through $\epsilon\propto s^{4/3}$. This is typically a very good approximation at the high temperatures achieved in the initial state of nucleus-nucleus collisions.
Our result is displayed in Fig.~\ref{fig:trento} as a shaded band. Note that we recombine 0.25\% bins into 1\% bins for sake of visualization.
Our \trento{} calculation is in good agreement with ATLAS data (open symbols) for both $\rho_2$ and $\rho_3$, and is consistent with the full hydrodynamic calculation shown in Fig.~\ref{fig:hydro}, in the sense that both evaluations slightly underestimates $\rho_2$ while they overestimates $\rho_3$.

Note that $\rho_2$ and $\rho_3$ measured by the ATLAS Collaboration have a slight dependence on the $p_t$ cut used in the analysis~\cite{Aad:2019fgl}. The difference between our results and experimental data is of the same order, or smaller, than the dependence of experimental results on the $p_t$ cuts. This feature is not captured by our prediction, which is independent of these cuts by construction. It would be therefore interesting to have new measurements of $\rho_n$ with a lower $p_t$ cut, of order $0.2$ or $0.3$~GeV/c, which is where the bulk of the produced particles sits. This may improve agreement between our evaluations and data.
It would also be interesting to have a full, high-statistics calculation of $\rho_n$ in hydrodynamics to study the deviations from the predictor Eq.~(\ref{rhoninitial}), and the dependence on the $p_t$ cut. 
Note that our initial-state predictor (\ref{rhoninitial}) might have a broader range of applicability than hydrodynamics itself\footnote{We thank the anonymous referee for suggesting this.}. 
It is already well known that the hypothesis that $v_n$ is proportional to $\varepsilon_n$ is more general than hydrodynamics~\cite{Heiselberg:1998es,He:2015hfa}. 
In the same way, our hypothesis that $\langle p_t\rangle$ is determined by the initial energy seems a natural consequence of conservation laws, and might still be valid when hydrodynamics is not. 

While the quantitative results shown in Fig.~\ref{fig:trento} depend on the parametrization of the \trento{} model, we show in Appendix~\ref{s:parameters} that the main qualitative features, for instance the fact that $\rho_n$ is positive in central collisions, are robust and model-independent. 
It is interesting though that the choice of parameters made here, namely, $p=0$, preferred from previous comparisons~\cite{Bernhard:2016tnd,Giacalone:2017uqx}, and $k=2$~\cite{Giacalone:2017dud}, also optimizes agreement with $\rho_n$ data.

Finally, we compare with results obtained with two different predictors of $\langle p_t\rangle$ introduced in Sec.~\ref{s:ei}: 
The initial size $R$~\cite{Bozek:2012fw}, and the entropy per unit area $S/A_e$~\cite{Schenke:2020uqq}. 
We thus evaluate $\rho_n$ by replacing $E_i/S$ with $1/R$\footnote{We use $1/R$ instead of $R$ because the correlation between $R$ and $\langle p_t\rangle$ is negative, as shown in Fig.~\ref{fig:ptpredictor} (a).} or $S/A_e$ in Eq.~(\ref{rhoninitial}). 
With $1/R$, both $\rho_2$ and $\rho_3$ are completely different (dashed lines in Fig.~\ref{fig:trento}).  
These results show that the correlation between $\langle p_t\rangle$ and $v_n^2$ is {\it not\/} driven by the event-by-event fluctuations of the fireball size. 
The predictor $S/A_e$ (dotted lines in Fig.~\ref{fig:trento}) gives results similar to $E_i/S$ for $\rho_2$, not for $\rho_3$. 
None of these alternative predictors allows one to reproduce the observed $\rho_2$ and $\rho_3$. 

\section{Conclusions}

We have shown that ATLAS results on $\rho_n$ in Pb+Pb collisions can be explained by hydrodynamics.  
The mechanism driving the correlation between the mean transverse momentum and anisotropic flow in Pb+Pb collisions can be traced back to the initial density profile, i.e., to the early stages of the collision. 
This implies in turn that this observable has limited sensitivity to the details of the hydrodynamic expansion in general, and to the transport coefficients of the fluid in particular, as nicely confirmed by the hydrodynamic results (Fig.~9) of Ref.~\cite{Schenke:2020uqq}. 
We have found that $\langle p_t\rangle$ fluctuations are driven by fluctuations of the initial energy over entropy ratio $E_i/S$, and not by the fluctuations of the fireball size as previously thought. 
By use of Eq.~(\ref{rhoninitial}), models of initial conditions that fit anisotropic flow data and multiplicity fluctuations also naturally reproduce the centrality dependence of $\rho_2$ and $\rho_3$ measured by the ATLAS Collaboration without any further adjustment. 
Note that experimental data are also available for p+Pb collisions, the study of which we leave for future work.

\section*{Acknowledgments}
FGG was supported by CNPq (Conselho Nacional de Desenvolvimento Cientifico) grant 312932/2018-9, by  INCT-FNA grant 464898/2014-5 and FAPESP grant 2018/24720-6.
G.G. and J.-Y.O. were supported by USP-COFECUB (grant Uc Ph 160-16, 2015/13). J.N.H. acknowledges the support of the Alfred P. Sloan Foundation, support from the US-DOE Nuclear Science Grant No.  de-sc0019175.
J.-Y. O. thanks Piotr Bo\.zek for discussions. We acknowledge useful discussions with Bj\"orn Schenke, Chun Shen, and Derek Teaney.

\appendix
\section{Varying the parametrization of the initial profile}
\label{s:parameters}

\begin{figure*}[t]
    \centering
    \includegraphics[width=.95\linewidth]{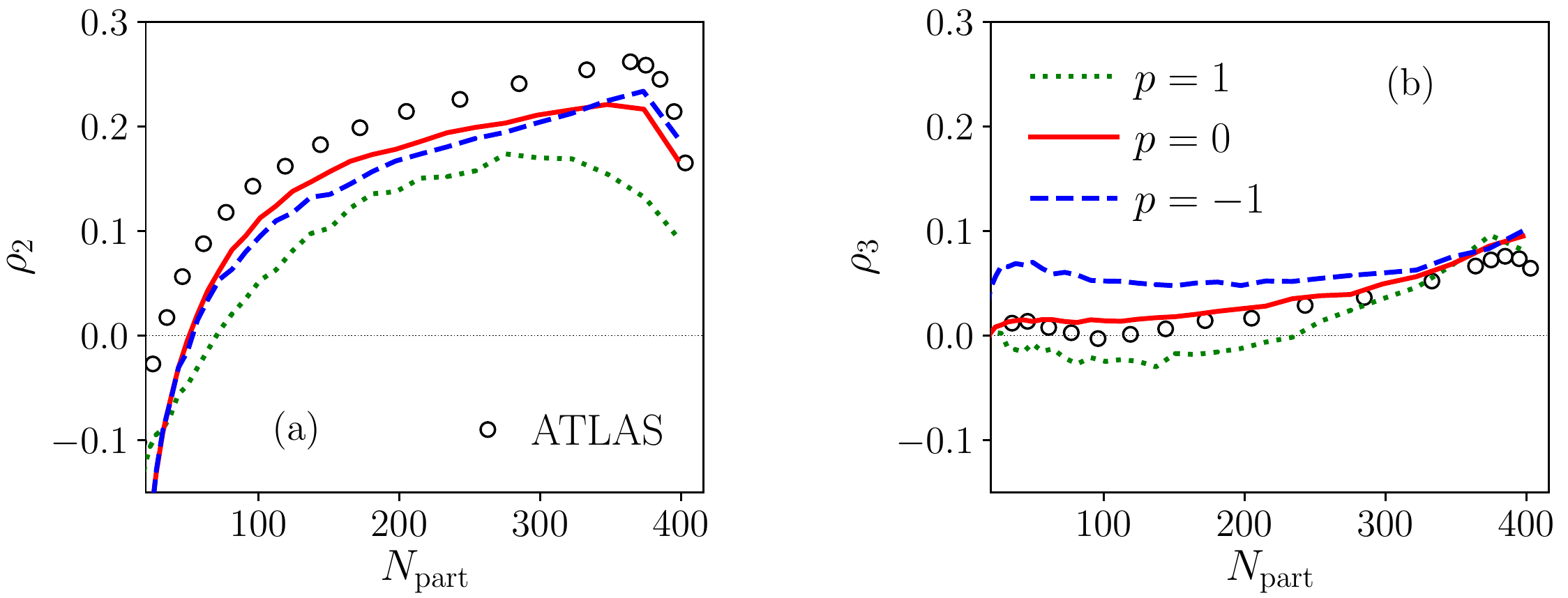}
    \caption{(Color online) Dependence of $\rho_2$ (a) and $\rho_3$ (b) on the parameter $p$ in the \trento{} model.  Full lines: $p=0$ (entropy density $s\propto \sqrt{T_AT_B}$), as in Fig.~\ref{fig:trento}. Dotted lines: $p=1$ ($s\propto T_A+T_B$). Dashed lines: $p=-1$ ($s\propto T_AT_B/(T_A+T_B)$).
We use broader centrality bins for this calculation than for Fig.~\ref{fig:trento}, which explains the small differences between the $p=0$ results of the two figures. 
      As in Figs.~\ref{fig:hydro} and \ref{fig:trento}, symbols are ATLAS data~\cite{Aad:2019fgl}.
    }
    \label{fig:trentop}
\end{figure*}

We check the sensitivity of $\rho_n$, as defined by Eq.~(\ref{rhoninitial}), to the parametrization of initial conditions.
Figure~\ref{fig:trentop} displays the variation of $\rho_n$ for three different values of $p$ in the \trento{} model.
Several qualitative trends are robust:  $\rho_2$ and $\rho_3$ are both positive for central collisions; 
As the number of participants decreases from its maximum value, $\rho_2$ steeply increases and then decreases, eventually becoming negative, while 
the centrality dependence of $\rho_3$ is milder.
But significant differences appear at the quantitative level, and the value $p=0$, which is the preferred value also for other observables~\cite{Bernhard:2016tnd,Giacalone:2017uqx}, agrees best with the recent $\rho_n$ data.
We have also studied the dependence on the parameter $k$ governing the magnitude of fluctuations in \trento{}. Results in Fig.~\ref{fig:trento} are obtained with $k=2$, but we have also carried out calculations with $k=1$, corresponding to larger fluctuations. We have found (not shown) that the results for $\rho_2$ are essentially unchanged except for a minor increase in central collisions, while the variation of $\rho_3$ becomes flatter, similar to the $p=-1$ results in Fig.~\ref{fig:trentop}. 

\end{document}